# Transport phenomena in sharply contrasting media with a diffusion barrier

O A Dvoretskaya[1] and P S Kondratenko[1,2]

[1]Nuclear Safety Institute, Russian Academy of Sciences, Bolshaya Tul'skaya St. 52, 115191 Moscow, Russia

[2]Moscow Institute of Physics and Technology (State University), Institutskii per. 9, Dolgoprudny, 141700 Moscow Region, Russia

E-mail: dvoriks@ibrae.ac.ru and kondrat@ibrae.ac.ru

**Abstract**

Using the advection-diffusion equation, we analytically study contaminant transport in a sharply contrasting medium with a diffusion barrier due to localization of a contaminant source in a low-permeability medium. Anomalous diffusion behavior and crossover between different transport regimes are observed. The diffusion barrier results in exponential attenuation of the source power, retardation of the contaminant plume growth, and modification of the concentration distribution at large distances.

## 1. Introduction

Anomalous diffusion is observed in systems which are far from equilibrium, in particular, it widely occurs in heterogeneous media [1]. Issues of great importance such as safe geoloical disposal of nuclear waste [2,3], membrane protein dynamics [4,5], conductivity of disordered semiconductors [6] are associated with it. It is commonly assumed that if the contaminant plume size depends on time according to the power law $R(t) \propto t^\gamma$ with $\gamma \neq \frac{1}{2}$, then anomalous diffusion (subdiffusion for $\gamma < \frac{1}{2}$ and superdiffusion for $\gamma > \frac{1}{2}$) takes place [1,7-9]. In addition, a crossover between different transport regimes (changing with time values of $\gamma$) can be observed [10-12]. Various approaches have been developed to describe non-classical transport such as the continuous time random walks (CTRW) [13] and aging

CTRW model that was first introduced in the context of diffusion in glasses [14], renormalization group method [15] initially devised within particle physics, fractional Fokker-Planck equation (FFPE) [16,17] and others. However, the above mentioned approaches should be used carefully. In particular, CTRW model describes the anomalous transport caused by trapping, which may have energetic, geometric or dynamical nature [18-20], but it can not be applied if the medium is not statistical homogeneous and FFPE is not valid for time-dependent forces [21]. To find out patterns of transport processes in heterogeneous media different models have been considered such as multi-length scale random fields [22], random velocity field [23,24], comb structure [25] and many others. The comb structure represents the regular heterogeneous sharply contrasting medium and can be studied using classical diffusion equation without any additional assumptions [26]. We paid a lot of attention to this kind of heterogeneous medium [26-30], which is treated, as two interacting subsystems of high and low permeability. General features of the transport phenomena in sharply contrasting media are the trapping of the particles by the low-permeable medium for long times, the subdiffusive dynamics, crossover between different transport regimes, multi-stage structure of the concentration distribution at large distances. Using the diffusion equation, contaminant transport caused by diffusion in the simplest realization of sharply contrasting media has been first studied by Dykhne et al. [27]. Later we generalized that model by introducing an advective term [29]. In the above mentioned works the contaminant source was located inside the high permeability region. The aim of the present paper is to analyze contaminant migration in sharply contrasting medium like [29] with a diffusion barrier due to the localization of the contaminant source inside the low-permeability region. We prove that the diffusion barrier significantly affects the transport processes. Also we show that the definition of the anomalous diffusion behavior mentioned in the beginning should be modified for problems with diffusion barrier.

The paper is organized as follows. In Sec. 2, we formulate the problem and derive some basic relations. In Sec 3, we analyze in detail the behavior of the contaminant concentration of the particles located inside the high-permeability region and study influence of the diffusion barrier on contaminant transport. We summarize our results in Sec. 4.

## 2. Problem formulation and basic relations

Let us consider contaminant transport in the heterogeneous system consisting of two parts (see Fig.1): a high-permeability medium I occupying a plane-parallel layer of the thickness $a$ (fracture) and a low-permeability medium II, filling the rest of the space (matrix). Transport in the fracture is provided by advection with velocity $\vec{u}$ and diffusion, whilst transport in the matrix is caused by diffusion only. The coordinate $z$ is chosen along the normal to the plane of the fracture $Oxy$ and the coordinate $x$ is along the advection velocity, so that $\vec{u}(z) = \begin{cases} \{u,0,0\} & \text{for } |z| < a/2 \\ 0 & \text{for } |z| > a/2 \end{cases}$. Let $\vec{\rho}$ be a two-dimensional radius vector $\vec{\rho} = \{x, y, 0\}$. The diffusivity is $D(z) = \begin{cases} D & \text{for } |z| < a/2 \\ d & \text{for } |z| > a/2 \end{cases}$, $D \gg d$. $c(\vec{\rho}, z; t)$ is the contaminant concentration distribution.

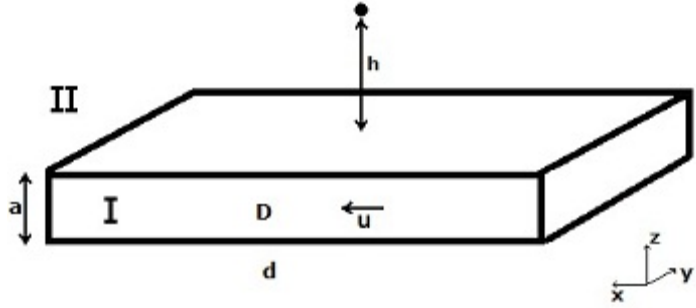

**Figure 1.** Regularly heterogeneous sharply contrasting medium

The advection-diffusion equation and boundary conditions have the following form

$$\frac{\partial c}{\partial t} + \nabla(\vec{u}(z)c) - div(D(z)\nabla c) = 0, \tag{1}$$

$$c(\vec{\rho}, z; t)\Big|_{z=\pm a/2-0}^{z=\pm a/2+0} = 0, \quad -D(z)\frac{\partial c(\vec{\rho}, z; t)}{\partial z}\Big|_{z=\pm a/2-0}^{z=\pm a/2+0} = 0. \tag{2}$$

The contaminant source is assumed to be located inside the matrix at distance $h$ from the fracture and given by the initial condition

$$c(\vec{\rho}, z; 0) = N_0 \delta(z - h - a/2)\delta(\vec{\rho}). \tag{3}$$

Taking the Fourier transform of equation (1) with respect to $\vec{\rho}$ and Laplace transform with respect to $t$, we obtain

$$\left(p + i\vec{u}\vec{k} + Dk^2 - D\frac{\partial^2}{\partial z^2}\right) c_{p\vec{k}} = 0 \quad \text{for } |z| < a/2, \tag{4}$$

$$\left(p + dk^2 - d\frac{\partial^2}{\partial z^2}\right) c_{p\vec{k}} = N_0 \delta(z - h - a/2) \quad \text{for } |z| > a/2, \tag{5}$$

where $p$, $\vec{k}$ is the Laplace and two-dimensional Fourier variables, respectively.

Further we call the particles located inside the fracture as active and our aim is to analyze their concentration distribution which is given by $n(\vec{\rho}; t) = \int_{-a/2}^{a/2} dz\, c(\vec{\rho}, z; t)$ for $t \gg \sqrt{t_h t_0}$, $t_0 = \frac{a^2}{4D}$.

By integrating (4) over $z$, we get

$$\left(Dk^2 + i\vec{u}\vec{k} + p\right) n_{p\vec{k}} + q_{p\vec{k}} = 0, \tag{6}$$

here $q_{p\vec{k}} = -D(z)\dfrac{\partial c_{pk}}{\partial z}\bigg|_{z=-a/2}^{z=a/2}$ is the Fourier-Laplace transform of the flux density.

To find $q_{p\vec{k}}$ we solve equation (5) with boundary conditions (2):

$$q_{p\vec{k}} = n_{p\vec{k}}\sqrt{\dfrac{p+dk^2}{t_1}} - N_0 \exp\left[-2\sqrt{t_h\left(p+dk^2\right)}\right], \qquad (7)$$

where $t_h = \dfrac{h^2}{4d}$, $t_1 = \dfrac{a^2}{4d}$.

Using (6), (7), we obtain active particles concentration in the Fourier-Laplace space

$$n_{p\vec{k}} = \dfrac{N_0 \exp\left[-2\sqrt{t_h\left(p+dk^2\right)}\right]}{Dk^2 + i\vec{u}\vec{k} + p + \sqrt{\dfrac{p+dk^2}{t_1}}}. \qquad (8)$$

To find out the influence of the diffusion barrier on the active particles transport, the comparison between the considered problem and the "barrier-free" one should be provided. Thus we solve the problem with the initial condition given by substitution $\delta(z-h-a/2) \to \delta(z)$ in equation (3), i.e. the source being located inside the fracture rather than in the matrix, that corresponds to the "barrier-free" problem studied in [29]. Hereafter, we denote the quantities related to this problem by the superscript $^*$. The active particles concentration per unit area $n^*_{p\vec{k}}$ can be found by solving equation (1) in the Fourier-Laplace space with the above modification of the initial condition. Thus we have

$$n^*_{p\vec{k}} = \dfrac{N_0}{Dk^2 + i\vec{u}\vec{k} + p + \sqrt{\dfrac{p+dk^2}{t_1}}}. \qquad (9)$$

Comparing this equation with (8), we find $n_{p\vec{k}} = n^*_{p\vec{k}} \exp\left[-2\sqrt{t_h\left(p+dk^2\right)}\right]$. We apply the inverse Fourier-Laplace transform to $n_{p\vec{k}}$ and use the change of variables $s = p+dk^2$. Thereby we obtain the relation

$$n(\vec{\rho},t) \cong \int_0^t dt' f(t') n^*(\vec{\rho}, t-t'), \qquad (10)$$

where

$$f(t) = \sqrt{\dfrac{t_h}{\pi t^3}} \exp\left(-\dfrac{t_h}{t}\right). \qquad (11)$$

and

$$n^*(\vec{\rho},t) \cong \int_{b-i\infty}^{b+i\infty} \frac{ds}{2\pi i} e^{st} \int \frac{d^2k}{2\pi} n^0_{s\vec{k}} \exp(-i\vec{k}\vec{\rho} - dk^2 t), \quad \text{Re } b > 0 \qquad (12)$$

with $n^0_{s\vec{k}} = \dfrac{N_0}{Dk^2 + i\vec{u}\vec{k} + s + \sqrt{\dfrac{s}{t_1}}}$.

Since the active particles concentration $n(\vec{\rho},t)$ is determined by integral convolution (10), we conclude that the problem with diffusion barrier is equivalent to the "barrier-free" one with the effective source locating inside the fracture and given by the function $N_0 f(t)$. It should be noted that above conclusion is extremely important for the considered problem. The obtained results have a clear physical interpretation in terms of the effective source.

Also we analyze key parameters of the contaminant transport such as the number of active particles

$$N(t) = \int d^2\rho \, n(\vec{\rho},t), \qquad (13)$$

and the size of contaminant plume

$$R^2(t) = \frac{1}{N(t)} \int d^2\rho \cdot n(\vec{\rho},t)(\vec{\rho} - \vec{X}(t))^2, \qquad (14)$$

where $\vec{X}(t)$ is the average displacement given by

$$\vec{X}(t) = \frac{1}{N(t)} \int d^2\rho \cdot n(\vec{\rho},t)\vec{\rho}. \qquad (15)$$

## 3. Results and discussion

*3.1. Concentration distribution behavior*

Let us analyze the concentration distribution given by (10). Taking into account that $f(t)$ is sharply peaked function and $\int_0^\infty dt\, f(t) = 1$, and using (10), for $t \gg t_h$ we have

$$n(\vec{\rho},t) \cong n^*(\vec{\rho},t). \qquad (16)$$

This is valid when $\rho$ is not too large so that $n^*(\vec{\rho},t')$ changes slightly at $t - t_h < t' < t$. As it was expected at $t \gg t_h$, the concentration distribution behaves mostly as in the barrier-free problem. So we mainly focus on the opposite case

$$t \ll t_h. \qquad (17)$$

For $t \ll t_h$ using the substitution $t - t'$ for $t'$ in (10) and Taylor series expansion, we get

$$f(t-t') \cong f(t)\exp\left(-\frac{t'}{t_{eff}}\right). \qquad (18)$$

with effective time

$$t_{eff} = \frac{t^2}{t_h}. \qquad (19)$$

Finally, using (10), (18), one obtains

$$n(\vec{\rho},t) = N_0 \sqrt{\frac{t_h}{\pi t^3}} \exp\left(-\frac{t_h}{t}\right) \cdot \int_0^\infty dt' \exp\left(-\frac{t'}{t_{eff}}\right) \cdot n^*(\vec{\rho},t'). \qquad (20)$$

This integral converges at $t' \sim t_{eff}$, so the effective source can be considered as continuously acting during the time interval equal $t_{eff}$.

### 3.2. Number of the active particles

The time dependence of the number of active particles $N(t)$ given by (13) can be obtained easily using the relation

$$N(t) = \int_{b-i\infty}^{b+i\infty} \frac{dp}{2\pi i} e^{pt} n_{p\vec{k}}\bigg|_{\vec{k}=0}. \qquad (21)$$

Hence, we have

$$N(t) = N_0 \int_{\sigma-i\infty}^{\sigma+i\infty} \frac{dp}{2\pi i} \frac{\exp\left(-2\sqrt{pt_h} + pt\right)}{p + \sqrt{p/t_1}}. \qquad (22)$$

This function is governed by the two factors: influence of the effective source, and diffusive particle flux from the fracture into the matrix. The number of active particles increases with time due to the first factor, and then it reaches the maximum at $t = 2t_h$ and decreases as $N(t) \propto \frac{1}{\sqrt{t}}$ because of the second factor.

Taking into account expression (22), we conclude that $\sqrt{t_1 t_h}$ is the time when the number of particles leaving the fracture equals the number of particles remaining in the fracture. In the "barrier-free" problem $t_1$ has the same meaning. For $t \ll \sqrt{t_1 t_h}$ the second term in denominator (22) should be neglected. After integration, we obtain

$$N(t) \cong \frac{N_0}{\sqrt{\pi}} \sqrt{\frac{t}{t_h}} \exp\left(-\frac{t_h}{t}\right), \quad t \ll t_h. \qquad (23)$$

For $t \gg \sqrt{t_1 t_h}$ the first term in denominator (22) is much less than the second one, so we get

$$N(t) \cong \frac{N_0}{\sqrt{\pi}} \sqrt{\frac{t_1}{t}} \exp\left(-\frac{t_h}{t}\right). \qquad (24)$$

In the "barrier-free" problem, the time dependence of the number of active particles $N^*(t)$ differs from the one given by (22) in the absence of the effective source term $\exp(-2\sqrt{pt_h})$. Thus we get

$$N^*(t) = N_0 \exp\left(\frac{t}{t_1}\right) \operatorname{erfc}\left(\sqrt{\frac{t}{t_1}}\right) \tag{25}$$

where the last factor is the complementary error function.

Figure 2 shows the $N(t)$ and $N^*(t)$ given by (22) and (25) respectively. They coincide at $t \gg t_h$ since the influence of the diffusion barrier becomes insignificant. Otherwise, it is clear that diffusion barrier causes renormalization of the source power.

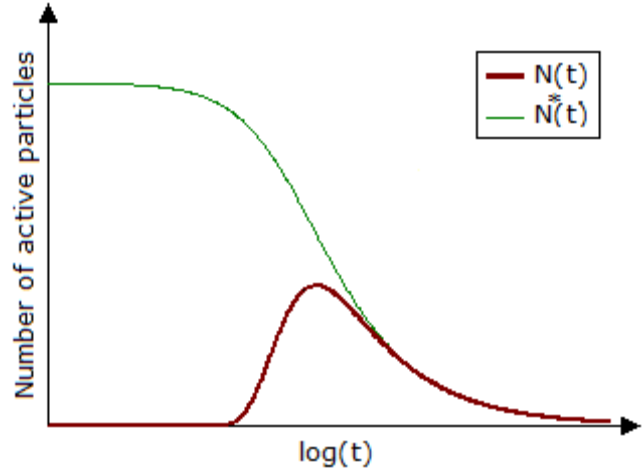

**Figure 2.** The evolution of the number of active particles in the problem with diffusion barrier and the "barrier-free" one.

### 3.3. Transport regimes

Transport regimes are characterized by the contaminant plume size $R(t)$ given by (14). We find both the plume size for the problems with and without diffusion barrier to detect the influence of the barrier on contaminant transport of active particles.

It is easy to find the contaminant plume size of the barrier-free problem by using the relation

$$\left(R^*(t)\right)^2 = -\frac{1}{N^*(t)} \int_{b-i\infty}^{b+i\infty} \frac{dp}{2\pi i} e^{pt} \left.\frac{\partial^2 n^*_{p\vec{k}}}{\partial \vec{k}^2}\right|_{\vec{k}=0} - \left(\vec{X}^*(t)\right)^2, \tag{26}$$

where the average displacement is given by

$$\vec{X}^*(t) = \frac{i}{N^*(t)} \int_{b-i\infty}^{b+i\infty} \frac{dp}{2\pi i} e^{pt} \left.\frac{\partial n^*_{p\vec{k}}}{\partial \vec{k}}\right|_{\vec{k}=0} \tag{27}$$

and $N^*(t)$ is defined by (21) with $n_{p\vec{k}} \to n^*_{p\vec{k}}$.

Let us consider the concentration given by (12). The inverse Fourier transform gives

$$n^*(\vec{\rho},t) = \frac{N_0}{4\pi dt} \int d^2\vec{\rho}\,' \exp\left[-\frac{(\vec{\rho}-\vec{\rho}\,')^2}{4dt}\right] n^0(\vec{\rho}\,',t) \tag{28}$$

The first exponential factor in the above equation is dominant at $t \gg t_2$, $t_2 = t_1\left(\frac{D}{d}\right)^2$ when the number of particles moving from the fracture to the matrix is so large that contaminant transport in the fracture is determined by matrix diffusion. Otherwise $t \ll t_2$, and the behavior of concentration is determined by $n^0(\vec{\rho},t)$.

There are several cases that differ from each other in the relations between the characteristic times $t_u$, $t_1$, $t_2$, where $t_u = \frac{4D}{u^2}$, thus we consider the most interesting ones: 1. $t_u \ll t_1 \ll t_2$, 2. $t_1 \ll t_u \ll t_2$.

1. $t_u \ll t_1 \ll t_2$

For $t \gg t_u$ the advection affects the contaminant transport in the fracture. The diffusive flux of particles into the matrix becomes significant at $t \gg t_3$, $t_3 = \left(t_u t_1^2\right)^{1/3}$, so the transport regimes are influenced by the interplay between this flux and advection. Using (27), we find the average displacement

$$\left|\vec{X}^*(t)\right| \cong \begin{cases} 0, & t \ll t_u \\ ut, & t_u \ll t \ll t_3 \\ \sqrt{4D_u t}, & t \gg t_3 \end{cases} \tag{29}$$

with $D_u = u^2 t_1$.

Using (29), we get the contaminant plume size given by (26)

$$R^*(t) \cong \begin{cases} \sqrt{4Dt}, & t \ll t_3 \\ \sqrt{4D_u t}, & t \gg t_3 \end{cases} \tag{30}$$

According to the definition given in the introduction $R^*(t) \sim t^\gamma$ with $\gamma = \frac{1}{2}$.

2. $t_1 \ll t_u \ll t_2$

Here, the advection doesn't affect the contaminant transport when $t \ll t_4$, $t_4 = \frac{t_u^2}{t_1}$ and the diffusive particles flux into the matrix acts at $t \gg t_1$. The average displacement and contaminant plume size take the following form

$$\left|\vec{X}^*(t)\right| \cong \begin{cases} 0, & t \ll t_4 \\ \sqrt{4D_u t}, & t \gg t_4 \end{cases} \tag{31}$$

$$R^*(t) \cong \begin{cases} \sqrt{4Dt}, & t \ll t_1 \\ \sqrt{4D\sqrt{t_1 t}}, & t_1 \ll t \ll t_4 \\ \sqrt{4D_u t}, & t \gg t_4 \end{cases} \tag{32}$$

It should be noted that for the first and third regimes we get $\gamma = \frac{1}{2}$ and for the second one - $\gamma = \frac{1}{4}$ that corresponds to subdiffusion.

Let us now turn to the problem with diffusion barrier. Taking into account (26), (27), and using (13)-(15), (19) one can obtain the relationship between the contaminant plume sizes in the problems with and without diffusion barrier

$$R(t) \sim \max\left\{ R^*(t_{eff}), \left|\vec{X}^*(t_{eff})\right| \right\} \tag{33}$$

$$\vec{X}(t) \sim \vec{X}^*(t_{eff}), \tag{34}$$

where $t_{eff}$ is given by (19).

Substituting $t$ with $t_{eff}$ into (29), (30), we obtain the average displacement and contaminant plume size for the problem with diffusion barrier in the case of $t_u \ll t_1 \ll t_h$, $t_2$. Similarly, using (31), (32), these quantities can be found in the second case of $t_1 \ll t_u \ll t_h$, $t_2$.

For $t_u \ll t_1 \ll t_h$, $t_2$, the contaminant plume size is

$$R(t) \sim \begin{cases} t\sqrt{\dfrac{D}{t_h}}, & t \ll \sqrt{t_u t_h} \\ u\dfrac{t^2}{t_h}, & \sqrt{t_u t_h} \ll t \ll \sqrt{t_3 t_h} \\ t\sqrt{\dfrac{D_u}{t_h}}, & \sqrt{t_3 t_h} \ll t \ll t_h \end{cases} \tag{35}$$

The average displacement $\left|\vec{X}(t)\right| = 0$ while $t \ll \sqrt{t_u t_h}$, otherwise $\left|\vec{X}(t)\right| \sim R(t)$.

For $t_1 \ll t_u \ll t_h$, $t_2$

$$R(t) \sim \begin{cases} t\sqrt{\dfrac{D}{t_h}}, & t \ll \sqrt{t_1 t_h} \\ \sqrt{Dt\sqrt{t_1/t_h}}, & \sqrt{t_1 t_h} \ll t \ll \sqrt{t_4 t_h} \\ t\sqrt{\dfrac{D_u}{t_h}}, & \sqrt{t_4 t_h} \ll t \ll t_h \end{cases} \tag{36}$$

The average displacement $\left|\vec{X}(t)\right| = 0$ when $t \ll \sqrt{t_4 t_h}$, otherwise $\left|\vec{X}(t)\right| \sim R(t)$.

Thus, we find the average displacement and contaminant plume size that determine transport regimes specific for the problem with diffusion barrier. All of the regimes given by (35), (36) have the retardation factor $t/t_h \ll 1$. Figure 3 shows the influence of this factor on contaminant plume size. It is clear that for $t \ll t_h$ the diffusion barrier results in retardation of the contaminant plume growth.

It is worth noting that $\gamma$ given by the relation $R(t) \sim t^\gamma$ with $R(t)$ from Eqs. (35),(36) does not determine the rate of the plume size growth because of the presence of the retardation factor $t/t_h \ll 1$. So transport regimes can not be identified based on these values of $\gamma$. Actually, $\gamma$ should be given by $R(t) \sim t_{eff}^\gamma$ in problems with diffusion barrier so that values of $\gamma$ coincide with the ones obtained in the "barrier-free" problem.

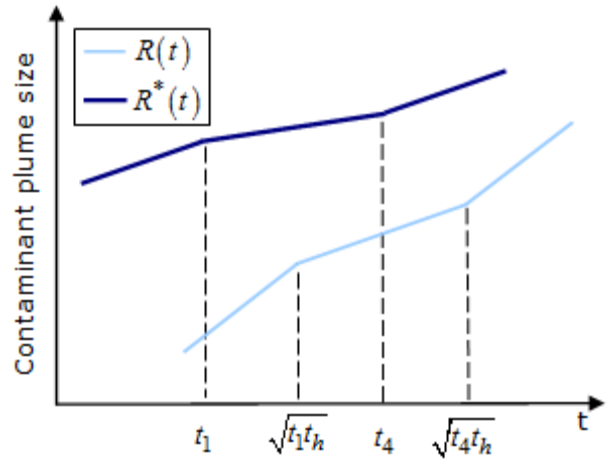

**Figure 3.** Log-Log scale plot of the contaminant plume size with respect to time for the problems with and without diffusion barrier. $R^*(t)$ and $R(t)$ are given by (32), (36) respectively.

Two remarks have to be made. The first one is related to the times $t \ll t_h \dfrac{d}{D} = \dfrac{h^2}{4D}$. Using (10) and (28), we find that the matrix diffusion governs the particles transport inside the fracture as long as the active particles number is extremely small. So for $t \ll t_h \dfrac{d}{D}$ the concentration of active particles takes the form

$$n(\vec{\rho},t) \propto \exp\left(-\frac{\rho^2 + h^2}{4dt}\right) \qquad (37)$$

The second remark is related to the times $t \ll \sqrt{t_0 t_h}$, where $t_0 = \dfrac{a^2}{4D}$. At these times the concentration distribution over $z$ is highly inhomogeneous. The above mentioned inequality can be obtained as follows. The diffusion length is $\delta z \sim \sqrt{D t_{eff}}$. The concentration is not homogeneous over $z$ if the diffusion length is much less than the fracture thickness $\delta z \ll a$, hence $\sqrt{Dt^2/t_h} \ll a$ and therefore $t \ll \sqrt{t_0 t_h}$. For these times, only two regimes occur if $t_h \frac{d}{D} \gg \sqrt{t_0 t_h}$: $R(t) \sim \sqrt{Dt^2/t_h}$, $R(t) \sim ut^2/t_h$ since $\sqrt{t_1 t_h} \gg \sqrt{t_0 t_h}$, otherwise $t_h \frac{d}{D} \ll \sqrt{t_0 t_h}$, the concentration distribution is given by (37).

*3.4. Concentration tails*

The diffusion barrier affects on the active particles transport at large distances so that the modification of concentration tails with respect to the found in the "barrier-free" problem [29] is observed.

First, we note that at large distances the function $n^*(\vec{\rho},t)$ from (20) can be represented as follows

$$n^*(\vec{\rho},t') \propto \exp\left(-\Phi^*(\vec{\rho},t')\right), \tag{38}$$

where $\Phi^*(\vec{\rho},t) \sim \left(\frac{|\vec{\rho}|}{R^*(t)}\right)^{\frac{1}{1-\gamma}}$ and $R^*(t) \sim t^\gamma$ given by Eqs. (30), (32).

For $\vec{\rho}$ satisfying the inequality $\Phi^*(\vec{\rho}, t_{eff}) \frac{t}{t_h} \ll 1$, the asymptotic concentration $n(\vec{\rho},t)$ takes a form

$$n(\vec{\rho},t) \propto \exp\left(-\Phi(\vec{\rho},t)\right), \tag{39}$$

with

$$\Phi(\vec{\rho},t) \cong \frac{t_h}{t} + \frac{\left(\Phi^*(\vec{\rho}, t_{eff})\right)^{1-\gamma}}{\gamma^\gamma (1-\gamma)^{1-\gamma}} \tag{40}$$

Otherwise $\Phi^*(\vec{\rho}, t_{eff}) \frac{t}{t_h} \gg 1$, using (10), (38) and then applying a Taylor series expansion for $\Phi^*(\vec{\rho}, t-t')$ around the point $t'=0$, one obtains

$$n(\vec{\rho},t) \propto \int_0^\infty \sqrt{\frac{t_h}{\pi t'^3}} \cdot \exp\left(-\frac{t_h}{t'} - \Phi^*(\vec{\rho},t) - t' \frac{\partial \Phi^*(\vec{\rho},t)}{\partial t}\right) dt'. \tag{41}$$

Taking advantage of the saddle-point method, we get

$$n(\vec{\rho},t) \propto \exp\left[-2\left(\frac{t_h}{t} \cdot \frac{\gamma}{1-\gamma} \Phi^*(\vec{\rho},t)\right)^{\frac{1}{2}} + \Phi^*(\vec{\rho},t)\right]. \tag{42}$$

For $t \ll t_h$ this expression describes the remote stages of the concentration tails as the first stage is given by (39). Comparing (38) with (39), (42), we conclude that the diffusion barrier result in the modification of concentration tails with respect to the found in the "barrier-free" problem. It follows from (41), (42) that the diffusion barrier affects the remote stages of the concentration tails even $t > t_h$

Now we analyze the concentration tails in the case $t_u \ll t_1 \ll t_h$. The concentration behavior at large distance is different for various time intervals. Further, we consider times $\sqrt{t_u t_h} \ll t \ll \sqrt{t_1 t_h}$ and $t \gg \sqrt{t_1 t_h}$, and show that the concentration tails have a multistage structure: they consist of several parts, for each of them the concentration behavior being different.

For $t \gg \sqrt{t_1 t_h}$ the first stage of the concentration tail defined by (39) corresponds to the distances $t\sqrt{D_u/t_h} \ll \rho \ll \sqrt{D_u t_h}$ and take a form

$$n(\vec{\rho},t) \propto \exp\left(-\frac{t_h}{t} - \frac{|\vec{\rho}|}{t\sqrt{D_u/t_h}}\right). \tag{43}$$

Remote stages of the tail given by (42) correspond to the distances $\sqrt{D_u t_h} \ll \rho \ll ut$, then $R^*(t) \cong \sqrt{4D_u t}$

$$n(\vec{\rho},t) \propto \exp\left(-\frac{\rho}{\sqrt{D_u t}}\sqrt{\frac{t_h}{t}} - \frac{\rho^2}{4D_u t}\right). \tag{44}$$

and $\rho \gg ut$, then $R^*(t) = \sqrt{4Dt}$

$$n(\vec{\rho},t) \propto \exp\left(-\frac{\rho}{\sqrt{Dt}}\sqrt{\frac{t_h}{t}} - \frac{\rho^2}{4Dt}\right). \tag{45}$$

Thus, for $t \gg \sqrt{t_1 t_h}$ the concentration tail consists of three stages given by (43), (44) and (45).

For $t \ll \sqrt{t_u t_h}$ the concentration at large distance can be found in a similar way. The case $\sqrt{t_u t_h} \ll t \ll \sqrt{t_1 t_h}$ deserves a special consideration.

For $\sqrt{t_u t_h} \ll t \ll \sqrt{t_1 t_h}$ using relation (10) and (38), we get

$$n(\vec{\rho},t) \propto \int_0^t dt' \exp\left[-\frac{t_h}{t-t'} - \frac{(\vec{\rho}-\vec{u}t')^2}{4Dt'}\right]. \tag{46}$$

It is necessary to analyze the concentration behavior at $x < ut$ and $x > ut$.

If $x < ut$

$$n(\vec{\rho},t) \propto \theta(x)\exp\left(-\frac{t_h}{t-x/u} - \frac{y^2}{4Dx/u}\right), \tag{47}$$

where $\theta(x)$ is the Heaviside step function.

The concentration in the region $x<0$ is exponentially smaller than in the region $x>0$, so using the step function in the above expression is reasonably.

Otherwise $x>ut$, taking advantage of the saddle-point method and using (46) we get the following expression

$$n(\vec{r},t) \propto \exp\{-\Phi(\vec{r},t)\},$$
$$\Phi(\vec{r},t) \cong \left[\frac{x^2-(ut)^2}{4Dt}\frac{t_h}{t}\right]^{\frac{1}{2}} + \frac{(\vec{\rho}-\vec{u}t)^2}{4Dt}; \quad x>ut. \quad (48)$$

## 4. Conclusions

Contaminant transport through a regularly heterogeneous sharply contrasting medium with a diffusion barrier has been studied analytically. The medium is assumed to be consisting of two interacting subsystems of a high and low permeability. It is worth noting that sharp contrast in properties of the geological media is typical and results in anomalous diffusion. The diffusion barrier emerges due to the localization of the contaminant source in the low-permeability medium (matrix) far enough from the high-permeability region (fracture).

The concentration distribution of the particles located inside the fracture (active particles) at short and large distances, the contaminant plume size and the number of active particles have been found. We have studied the influence of the diffusion barrier on the contaminant transport by the comparing the results obtained for the problems with and without diffusion barrier.

It has been shown that at $t \ll t_h$ the problem with diffusion barrier is equivalent to the "barrier-free" one with the effective source locating inside the fracture and continuously acting during the effective time $t_{eff} = t^2/t_h$, otherwise $t \gg t_h$ - with the instant source locating in the fracture too. For $t \gg t_h$ the diffusion barrier results in the modification of the concentration tails only. So we mainly focus on times $t \ll t_h$ when the influence of the diffusion barrier on the active particles transport is extremely significant. Depending on the time interval, we have observed one of the several transport regimes specific for the diffusion barrier problem: $R(t) \sim ut^2/t_h$, $R(t) \sim t\sqrt{D/t_h}$, $R(t) \sim t\sqrt{D_u/t_h}$, $R(t) \sim \sqrt{Dt\sqrt{t_1/t_h}}$, where $R(t)$ is the contaminant plume size given by (14). Note that all the regimes have a retardation factor of $t/t_h \ll 1$, i.e. the contaminant plume grows slowly than in the "barrier-free" problem [29]. Therefore, we should use the relation $R(t) \sim t_{eff}^{\gamma}$ instead of $R(t) \sim t^{\gamma}$ to find the value of $\gamma$ and identify transport regime in problems with diffusion barrier; otherwise we obtain the physical meaningless results. In above mentioned expressions for $R(t)$ $\gamma$ equals 1, $\frac{1}{2}$ and $\frac{1}{4}$ respectively. So we have the modified by diffusion barrier advection, diffusion, quasidiffusion and subdiffusion.

The concentration distribution at large distances (concentration tail) has an exponential form and a multistage structure as in our previous works [24,26-30]. A modification of the concentration tails relative to the "barrier-free" problem is observed.

Evolution of the number of active particles has been found: $N(t) \cong \frac{N_0}{\sqrt{\pi}} \sqrt{\frac{t}{t_h}} \exp\left(-\frac{t_h}{t}\right)$ for $t \ll \sqrt{t_1 t_h}$ and $N(t) \cong \frac{N_0}{\sqrt{\pi}} \sqrt{\frac{t_1}{t}} \exp\left(-\frac{t_h}{t}\right)$ for $t \gg \sqrt{t_1 t_h}$. In the latter expression the prefactor is caused by the diffusive flux of the active particles from the fracture, while the exponential factor is governed by the effective.

The main results can be summarized as follows. The diffusion barrier significantly affects the active particles transport and results in 1. renormalization of the source power, namely, exponential attenuation of the power; 2. retardation of the contaminant plume growth; 3. modification of concentration tails.

## Acknowledgments

We acknowledge support from the Russian Foundation for Basic Research (RFBR) under Project 09-08-00573a, and support from the Federal Target Program "Scientific and academic staff of innovative Russia" for the period 2009-2013 under Contract No. 02.740.11.0746, and support Savannah River Nuclear Solutions, LLC (USA) under Subcontract No. AC81967N.